\documentclass[intlimits,twoside,a4paper]{article}

\usepackage{amsmath,amssymb}
\usepackage{graphicx}

\usepackage[T2A]{fontenc}
\usepackage[cp1251]{inputenc}

\usepackage{cmpj2}

\issue{2014}{17}{1}{13701}
\doinumber{10.5488/CMP.17.13701}

\title[Plasmon resonance sensitivity in fine metal particle]%
{Plasmon resonance sensitivity in fine metal particle}
\author[N.I. Grigorchuk]{N.I. Grigorchuk}
\address{
Bogolyubov Institute for
Theoretical Physics, National Academy of Sciences of Ukraine, \\
 14-b Metrologichna St., 03680 Kyiv, Ukraine}
\date{Received September 26, 2013, in final form November 4, 2013}
\authorcopyright{N.I. Grigorchuk, 2014}

\begin{document}

\maketitle

\begin{abstract}
The sensitivity of the wavelength position of surface plasmon resonances
in  prolate and oblate fine metal particles to the refractive index of an embedding solution,
a particle shape, and electron temperature is studied theoretically in the framework of different
analytical methods. All calculations are illustrated on the single potassium nanoparticle as an example.
\keywords sensitivity, surface plasmon resonance, fine metal particles
\pacs {78.67.-n}, {65.80.-g}, {73.23.-b}, {68.49.Jk}, {52.25.Os}
\end{abstract}

The wavelength, intensity, and shift of the surface plasmon resonance
(SPR) \cite{M} of fine metal particles (MPs) are sensitive to changes in
different physical parameters. The possibility to measure the optical effects with
the refractive index variation has been often used to detect changes in the surrounding
medium for chemical sensing or for probing the dynamics of biological molecules \cite{MLL}.
An important group of optical sensors is based on exploiting SPR to detect small
refractive index changes in close proximity to the sensing surface of MPs
\cite{LeS,BTJ,ML,TPF,MSS,HZS,YXW,CWG,G} and are associated with a biomolecule sensing \cite{HZS,YXW}.
Several authors have demonstrated, on the example of noble nanoparticles, how the SPR
position depends on the environment and the morphology of nanoparticles \cite{LeS,BTJ,ML,MSS}.
Much less attention has been paid to the study of sensitivity modes
of a spheroidal MP with a small effect of a core polarization, where the
contribution of the bound electrons can be neglected (e.g., K, Na, Be, Al, etc).

In this Letter, using the new approach specifying the nanoparticle shape,
we propose the expressions for calculation of the plasmonic sensitivity to
the changes in the main factors that significantly affect the position of the SPR
modes on the example of potassium nanoparticle.

1. The MPs considered are small in size relative to the wavelength of light.
If we change the refractive index of the embedding me\-di\-um $n$ by a definite
value, $\delta n$,  the sensitivity can be expressed in frequency units
as $S_{j,\omega}(n)=\partial \omega_{j,\mathrm{res}}/\partial n$. The resonance frequency
(or the corresponding wavelength) for nonspherical MP is given by \cite{BTJ}
\begin{equation}
 \label{eq om}
  \omega_{j, \textrm{res}} =
   \omega_\textrm{pl}\Big/\sqrt{\varepsilon_{\infty}+(1/L_j-1)n^2}\, ,
    \end{equation}
\begin{equation}
 \label{eq lm}
  \lambda_{j, \textrm{res}} = \lambda_\textrm{pl}\sqrt{\varepsilon_{\infty}+(1/L_j-1)n^2}\,,
   \end{equation}
where the Drude model with $\textrm{Re}(\varepsilon) \simeq \varepsilon_{\infty}-\omega^2_\textrm{pl}/\omega^2$
is used, $\omega_\textrm{pl}$ is the plasma frequency, $\varepsilon_{\infty}$ is the high frequency
dielectric constant, and $L_j$ refers to the Osborn's demagnetizing factors \cite{Osb} in $j$-th
direction.

We have restricted ourselves to the MPs with a sphero\-idal shape.
Then, variables $L_{\|}$ and $L_{\bot}$ will play the roles of
longitudinal (directed along the revolution axis of a spheroid) and
transverse (directed across this axis) components of the $L$ factor.
Since Osborn's factors are somewhat cumbersome to be used directly,
they can be changed with a high accuracy by the following relations
\begin{equation}
 \label{eq el}
  L_{\|} = \frac{1}{(1+R_{\|}/R_{\bot})^z}\,,
   \qquad L_{\bot} = \frac{1}{2}\left(1-L_{\|}\right),
    \end{equation}
where $z=\log{3}/\log{2}$, $R_{\|}$ and $R_{\bot}$ are the spheroid semiaxes
directed along and across the revolution axis of a spheroid, correspondingly.

Equation (\ref{eq lm}) describes the change in the SPR wavelength when the
factors $L_j$ or refractive index are changed. Figure~\ref{fig1} depicts the result of
calculations for a \emph{rodlike} potassium nanoparticle. One can see that the SPR peak wavelength
$\lambda_{\rm res}$ (normalized by plasma wavelength $\lambda_{\rm pl} = 2822$~\AA{}
for K) linearly depends on $n$, and is shifted to the red spectral side as $n$
is increased. Physically, this implies that the Cou\-lomb attractive force between
an electronic cloud and a positive metal lattice is weakened with $n$ and the
energy of SPR is reduced. The linear increase of SPR peak with $n$ retains for
different values of the aspect ratio $R_{\|}/R_{\bot}(\equiv a/b = x)$. Only the
line slope changes: it enhances for MPs with larger aspect ratios. From the slope
of the linear fit, the refractive index sensitivity $\partial\lambda_\textrm{res}/\partial n$
was calculated to be 256 nm per refractive index unit (RIU) for $S_{\bot}$ and 380 nm
per RIU for $S_{\|}$, with ``golden'' $x = 1.618$, or to be 205~nm/RIU
for $S_{\bot}$ and 2100~nm/RIU for $S_{\|}$, with $x = 16.18$.
A tenfold increase of the potassium particle aspect ratio enhances the $S_{\|}$-sensitivity
to the changes in the refractive index roughly 5 times.
Higher spheroid aspect ratio supports both the $\|$-SPR mode at
a longer wavelength and the $\bot$-mode at a rather shorter wavelength.
One can see that the $\|$-plasmon mode is much more sensitive
to the change in the refractive index than the transverse mode.
\begin{figure}[htb]
\centerline{
\includegraphics[width=0.6\textwidth]{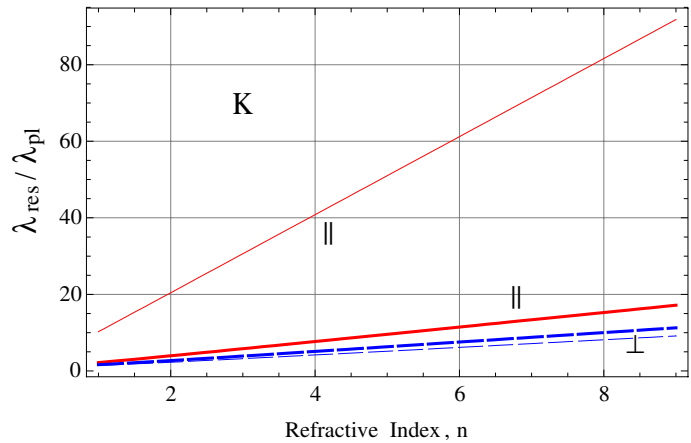}
}
\caption{(Color online) SPR peak wavelength modes ($\|$ and $\bot$) position as a function
of medium refractive index for prolate potassium nanoparticle with different aspect ratios:
$R_{\|}/R_{\bot}$= 1.618 (thick lines) and = 16.18 (thin lines).
\label{fig1}}
\end{figure}

2. By an analogy with the previous case, the sensitivity to the changes in the aspect
ratio of a spheroidal nanoparticle $x$ can be expressed in frequency units as
$S_{j,\omega}(x)=\partial \omega_{j,\textrm{res}}/\partial x$. Using equations~(\ref{eq om}) and (\ref{eq lm}) in the determination of sensitivity, one obtains
\begin{equation}
 \label{eq eso}
  S_{j,\omega}(x) = \frac{n\omega_\textrm{pl}[1/L_j(x)-1]}
  {\left\{\varepsilon_{\infty}+n^2[1/L_j(x)-1]\right\}^{3/2}}\,,
   \end{equation}
or
\begin{equation}
 \label{eq esl}
  S_{j,\lambda}(x)=\frac{n \lambda_{\rm pl}[1/L_j(x)-1]}{\sqrt{\varepsilon_{\infty}+n^2[1/L_j(x)-1]}}\,.
   \end{equation}
We can see that the sensitivity of the SPR to the medium can be enhanced by tuning the morphology
of the nanoparticle \cite{LeS,BTJ,ML,MSS}. The shape factor $L_j(x)$, which governs the geometric
tunability of the SPR, also determines the plasmon sensitivity of nanoparticles. It is easy
to find those values of $L_j$ for which the sensitivity $S_j(x)$ becomes maximal.
They can be expressed in the wavelength or frequency units by means of expressions
\begin{equation}
 \label{eq els}
  L_{\|,\lambda} = \frac{n^2}{n^2-2\varepsilon_{\infty}}\,,
   \qquad   L_{\|,\omega} = \frac{n^2}{n^2+2\varepsilon_{\infty}}
    \end{equation}
for a maximal longitudinal component of the $S_{\|}$, or
\begin{equation}
 \label{eq elo}
  L_{\bot,\omega} = \frac{2\varepsilon_{\infty}-n^2}{2\varepsilon_{\infty}+n^2}
   \end{equation}
for a maximal transverse component of the $S_{\bot}$.

Only a few MPs and dielectric media conform to the first of the two equation~(\ref{eq els}).
For this reason, we propose somewhat different approach. The derivative
$\partial\lambda_\textrm{res}/\partial n$ can be represented as a derivative of a composite function
\begin{equation}
 \label{eq cf_0}
  \frac{\partial\lambda_\textrm{res}}{\partial n} = \frac{\partial\lambda_\textrm{res}}{\partial\varepsilon'}
   \frac{\partial \varepsilon'}{\partial n}\,,
   \end{equation}
where $\varepsilon'$ is the real part of a particle dielectric function.
In the high frequency limit of dielectric function, $\lambda_\textrm{res}$ is
\begin{equation}
 \label{eq lmb}
  \lambda_\textrm{res} = \lambda_\textrm{pl} \sqrt{\varepsilon_{\infty}-\varepsilon'}\,.
   \end{equation}
From the plasmon resonance condition
\begin{equation}
 \label{eq ve}
  \varepsilon' = -(1/L_j-1) n^2,
   \end{equation}
and finally, we get
\begin{equation}
 \label{eq cf}
  \frac{\partial\lambda_{j,\textrm{res}}}{\partial n} = \frac{\lambda^2_\textrm{pl}}{\lambda_{j,\textrm{res}}} (1/L_j-1)n.
   \end{equation}

In the case of MP having a spherical form, the factor $L_j$
should be put equal to 1/3 in all the above equations.

3. Some experimental studies \cite{LeS,BTJ,ML,CWG,JaS,SRJ,CKY,KCZ} were carried
out on metal nanoparticles with different geometry to find the best nanoparticle
configuration to enhance the sensitivity of the SPR response.
Let us theoretically find the best geometrical factors $L_j$ to enhance
the sensitivity of the SPR response for MPs with a spheroidal shape.
To define the spheroid shape which provides the maximal sensitivity,
one can use the following relations:
\begin{equation}
 \label{eq erl}
  \left(R_{\|}/R_{\bot}\right)_\textrm{long} =
   \left(L_{\|,\omega}\right)^{-1/z}-1,
    \end{equation}
\begin{equation}
 \label{eq ert}
  \left(R_{\|}/R_{\bot}\right)_\textrm{trans} =
   \left(L_{\bot,\omega}\right)^{-1/z}-1,
    \end{equation}
for longitudinal and transverse components, correspondingly.

\begin{figure}[ht]
\centerline{
\includegraphics[width=0.6\textwidth]{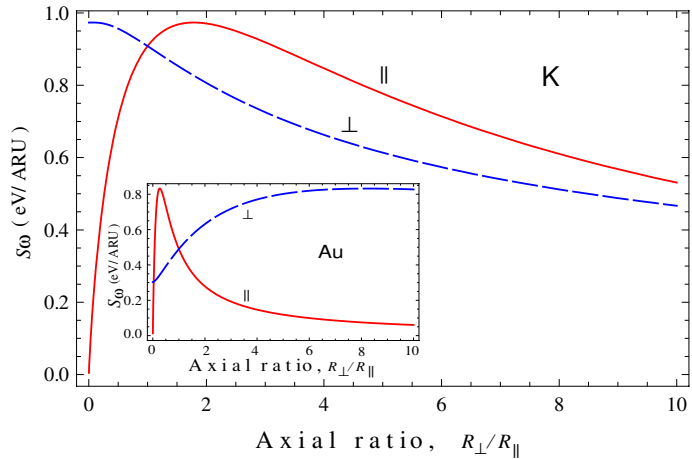}
}
\caption{
\footnotesize (Color online) Dependence of the sensitivity of the plasmon
resonance components (longitudinal~--- solid line and transverse~--- dashed line)
vs semiaxial ratio for spheroidal potassium nanoparticle
embedded in the SiO$_2$ matrix ($n = 1.56$). The inset shows the same dependence
for Au nanoparticle embedded in the water.
\label{fig2}}
\end{figure}
For instance, if we take the Au nanoparticle with $\varepsilon_{\infty}=9.84$ \cite{CWG}
embedded in water ($n=1.33$), then from the second equation~(\ref{eq els}) and equation~(\ref{eq elo}),
we obtain $L_{\omega}\simeq 0.082$ for $\|$ and $L_{\omega}\simeq 0.835$ for $\bot$ components,
which enable us to get from equations~(\ref{eq erl}) and (\ref{eq ert}) the maxima at
$(R_{\|}/R_{\bot})_\textrm{long} = 3.76$ and at $(R_{\|}/R_{\bot})_\textrm{trans} = 0.12$.

Figure~\ref{fig2} shows the dependence of plasmon sensitivity components on the axial
ratio (in axial ratio units (ARU)) calculated for the potassium nanoparticle with
$\varepsilon_{\infty}=1.24$ \cite{K}. We have found that the maximal sensitivity
can be realized at the axial ratio $(R_{\bot}/R_{\|})_\textrm{long}\simeq 2$ for
$S_{\|}$, and at $(R_{\bot}/R_{\|})_\textrm{trans}\simeq 0$ for $S_{\bot}$. The inset
in figure~\ref{fig2} displays the same dependence $S_E(x)$ for Au nanoparticle. Maximal
sensitivities are attained for Au at $(R_{\bot}/R_{\|})_\textrm{long}\simeq 0.26$ and
$(R_{\bot}/R_{\|})_\textrm{trans}\simeq 8.5$ for $S_{\|}$ and $S_{\bot}$, correspondingly.
Such quantities of maximal sensitivities for potassium and Au nanoparticles
can be obtained from equations~(\ref{eq erl}) and (\ref{eq ert}) as well.
The data in figure~\ref{fig2} for \emph{prolate} and \emph{oblate} potassium and Au nanoparticles
display a clear evidence of the spectral shift of the both components
$S_{\|}$ and $S_{\bot}$ as a function of spheroidal axial ratios.
Such a behavior of both $S(x)$ components is far from the linear one
predicted, e.g., in \cite{LeS} for the single Au nanorod.
\begin{figure}[htb]
\centerline{
\includegraphics[width=0.6\textwidth]{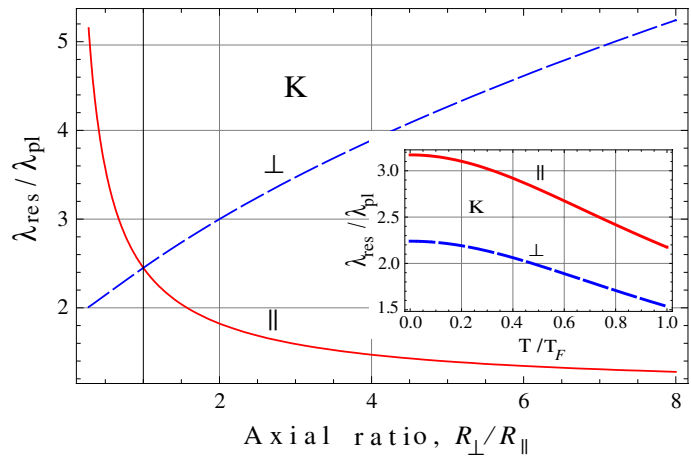}
}
\caption{
(Color online) The SPR peak
wavelength components ($\|$ and $\bot$) as a function of the spheroid
axial ratio for potassium nanoparticle embedded in the SiO$_2$ matrix. The inset shows
$\lambda_\textrm{res}(T)$ for potassium nanoparticle (with $x=1.62$) embedded in the SiO$_2$ matrix.
For potassium the Fermi temperature is $T_\textrm{F}=2.46\times 10^4$~K \cite{K}
and electron concentration is $n_0$=1.4$\times 10^{22}$ \cite{K}.
\label{fig3}}
\end{figure}

In figure~\ref{fig3} the dependence of the ratio $\lambda_\textrm{res}/\lambda_\textrm{pl}$ versus spheroid
axial ratio is depicted. As it is seen in this figure, when the aspect ratio for
\emph{rodlike} potassium nanoparticle increases, the $\|$-component of the
peak of the SPR wavelength $\lambda_\textrm{res}$ exhibits a blue shift,
whereas its $\bot$-component exhibits a red shift. For \emph{oblate}
potassium nanoparticle, vice versa, with an increase of the semiaxial ratio $R_{\bot}/R_{\|}$,
the $\|$-component of the SPR wavelength shifts to the red side,
whereas its $\bot$-component shifts to the blue side. From the slope
of the linear dependence of $\lambda_{\bot,\textrm{res}}$ on the axial ratio,
the shape sensitivity $\partial\lambda_\textrm{res}/\partial x$ becomes
roughly as high as 84~nm/ARU for $S_{\bot}$.

4. To characterize the thermal stability of the photonic crystal devices,
the study of their thermal sensitivity is very important \cite{FSK,YBA,DM}. It can be
expressed in wavelength units as $S_{j,\lambda}(T)=\partial \lambda_{j,\textrm{res}}/\partial T$.
The temperature dependence of $\lambda_\textrm{res}(T)$ comes from the
$T$-dependence of the electron concentration inside the MP \cite{L}
\begin{equation}
 \label{eq ect}
  n_\textrm{e}(T) = n_0 \left[1+\frac{\pi^2}{8}\left(\frac{k_\textrm{B} T}{\mu_0}\right)^2\right],
   \end{equation}
where $\mu_0=\mu(T=0)$ is chemical potential at zero temperature and $n_0=(2m\mu_0)^{3/2}/3\pi^2\hbar^3$.
Combining these equations with $\lambda_\textrm{pl}(T)=2\pi c/\sqrt{4\pi n_\textrm{e}(T) e^2/m}$
and equation~(\ref{eq lm}), we obtain
\begin{equation}
 \label{eq lmt}
  \lambda_{j, \textrm{res}}(T) \simeq \lambda_{\rm pl}(0)\sqrt{\frac{\varepsilon_{
   \infty}+(1/L_j-1)n^2}{1+({\pi^2}/{8})\left({k_\textrm{B} T}/{\mu_0}\right)^2}}\,.
    \end{equation}
Therefore,
\begin{equation}
 \label{eq lms}
  S_{j,\lambda}(T) \simeq \lambda_{\rm pl}(0)\frac{\pi^2}{8}
   \frac{\sqrt{\varepsilon_{\infty}+(1/L_j-1)n^2} }{\left[1+
    ({\pi^2}/{8})\left({k_\textrm{B} T}/{\mu_0}\right)^2\right]^{3/2}}
     \left(\frac{k_\textrm{B}}{\mu_0}\right)^2 T.
     \end{equation}
Equations (\ref{eq lmt}) and (\ref{eq lms}) allow one to estimate the shift of SPR
wavelength position with temperature and the temperature sensitivity for any ellipsoidal MPs.
It is easy to see from equation~(\ref{eq lmt}) and from the inset in figure~\ref{fig3} that
the value of $\lambda_\textrm{res}$ is shifted with $T$ to the shorter wavelength side,
which agrees well with the results of works \cite{DM, YBA}.

In summary, we propose the formulae that enable one to evaluate the plasmonic sensitivity
to the changes in the main factors considerably affected the SPR mode position in MPs.
For prolate (oblate) potassium and Au nanoparticles, there was found a nonlinear behavior
of both sensitivity components ($S_{\|}$ and $S_{\bot}$) as a function of an axial ratio.
The effect of the temperature variation on the plasmon sensitivity was demonstrated
on the example of a potassium nanoparticle.

%
%
\newpage
\ukrainianpart

\title{Чутливість плазмонних резонансів у малих металевих частинках}
\author{М.І. Григорчук}
\address{
Інститут теоретичної фізики ім.~М.М.~Боголюбова НАН України, \\
 вул.~Метрологічна, 14-б, 03680 Київ, Україна}

\makeukrtitle

\begin{abstract}
\tolerance=3000%
В рамках різних аналітичних методів теоретично вивчається чутливість положення довжин хвилі
резонансів поверхневого плазмона у витягнутих і сплюснутих малих металевих частинках до
показника заломлення оточуючого розчину, форми самої частинки та електронної температури.
Всі обчислення проілюстровані на прикладі окремої наночастинки калію.
\keywords чутливість, поверхневий плазмонний резонанс, малі металеві частинки

\end{abstract}

\end{document}